# Effect of Initial Disturbance on The Detonation Front Structure of a Narrow Duct

Hua-Shu Dou[*], Boo Cheong Khoo,

Department of Mechanical Engineering, National University of Singapore,
Singapore 119260

*tsldh@nus.edu.sg; huashudou@yahoo.com

**Abstract**

The effect of an initial disturbance on the detonation front structure in a narrow duct is studied by three-dimensional numerical simulation. The numerical method used includes a high resolution fifth-order weighted essentially non-oscillatory scheme for spatial discretization, coupled with a third order total variation diminishing Runge-Kutta time stepping method. Two types of disturbances are used for the initial perturbation. One is a random disturbance which is imposed on the whole area of the detonation front, and the other is a symmetrical disturbance imposed within a band along the diagonal direction on the front. The results show that the two types of disturbances lead to different processes. For the random disturbance, the detonation front evolves into a stable spinning detonation. For the symmetrical diagonal disturbance, the detonation front displays a diagonal pattern at an early stage, but this pattern is unstable. It breaks down after a short while and it finally evolves into a spinning detonation. The spinning detonation structure ultimately formed due to the two types of disturbances is the same. This means that spinning detonation is the most stable mode for the simulated narrow duct. Therefore, in a narrow duct, triggering a spinning detonation can be an effective way to produce a stable detonation as well as to speed up the deflagration to detonation transition process.

## 1. Introduction

The cellular detonation structure has attracted the interest of many researchers [1-5]. Although the phenomenon of detonation has been described for over 80 years, understanding of the structure and of the mechanisms involved has not yet been fully clarified. Detonation denotes the process of combustion in which a shock wave propagating at supersonic speed into unburned fuel mixture precedes the reaction zone. In the shock wave, unburned material is compressed and temperature is increased, resulting in a large increase in the reaction rate. The energy release from combustion behind the shock wave results in expansion which in its turn supports the shock wave, resulting in a self-supported propagating structure known as a detonation wave. This contrasts with a second model, in which a combustion wave propagates at subsonic speed, which is commonly referred as a deflagration wave. Most combustion processes start as a deflagration wave due to the relatively low input ignition energy required. However, this deflagration wave can turn into a detonation wave, in a process known as deflagration to detonation transition (DDT). In the DDT process, a disturbance imposed on the flow may have an important effect un the formation and the structure of the detonation wave. Likewise, a better understanding of the physics of detonation and how it appears has direct implications for industrial safety and damage prevention in chemical processes and in civil engineering. This provides the motivation for the present study on the influence of initial disturbances on the development and evolution of propagating detonation waves.

In recent years, numerical simulation has proved to be an effective tool in the study of the behaviour and the structure of detonation waves. Numerous simulations for two-dimensional (2D) duct [6-13], 2D variable cross section channel [14-16], and even three-dimensional (3D) ducts [17-23] have been conducted. These simulation results have helped providing a better understanding of the mechanism of detonation propagation under various conditions. It has become clear that numerical simulation is an effective tool for exploring the behaviour and the structure of detonation waves, hence helping to understand the physics of detonation structure under different conditions.

Both numerical simulations and experimental results have shown that there are three main types of cellular detonation structures, namely, rectangular, diagonal, and spinning modes [17-25]. Williams et al. [17] first observed what Hanana et al. [23] identify as a "rectangular structure" in their 3D simulations of detonation front in a rectangular duct, using a single Arrhenius kinetic model. Hanana et al [24] performed experiments in which they could obtain both rectangular and diagonal structures, and as a result they proposed these two expressions to identify them. Subsequently, Tsuboi et al. [18] obtained both detonation structures in simulations using a complex kinetic scheme, which confirmed the earlier observations. Hanana et al. [24] showed that the rectangular structure displays orthogonal waves traveling independently of each other on the four walls, with the triple point lines thus moving parallel to the opposite walls. Soot records in this case show the classical diamond cell patterns, including the presence of "slapping waves." For the case where the transverse waves move along the diagonal line of the duct cross section, diagonal front structures are formed with the triple point formation on a plane normal to the diagonals. Here the axes of the transverse waves are slanted at approximately 45° with respect to the wall which account for the lack of "slapping waves" on the diagonal structures. Pressure records indicate that where the intensity of the shock front is higher, the averaged wave velocity is larger, and the length of the detonation cell is shorter in forming the diagonal structures. It was suggested that the rectangular mode is effectively a superposition of two-dimensional orthogonal structures, while the diagonal structures are fundamentally three-dimensional [24]. In their experiments [24], the detonation is first ignited in a high pressure section, and then it transits to a low pressure section through a Mylar diaphragm. The final pattern recorded on the downstream wall is controlled by modulating the way of the diaphragm breaking. It was concluded that the way the detonation ignition occurs is the key parameter to induce one or the other type of structure. However, the relationship (if any) between the two different types of structures is not completely clarified, and the formation mechanism of the structure is still not clear.

In the simulation work of Tsuboi et al. [18], the rectangular mode is further sub-divided into two types, namely, in phase and out of phase. Their subsequent study on 3D simulation of detonation in both circular and rectangular ducts shows the presence of a spinning detonation front [25].

Deledicque and Papalexandris [22] also carried out 3D simulation of detonation in a rectangular duct. They employed the one-step Arrhenius chemical reaction model and the influence of geometry was considered for three parameter sets. Two different types of boundary conditions namely periodic and reflection boundary conditions were imposed on the walls. The above-mentioned two detonation structures of rectangular and diagonal modes were captured by suitably perturbing the initial conditions. It was also found that for the "in-phase" rectangular and diagonal structures, there is similarity in the geometrical evolution of the front. Likewise, the difference in cell lengths observed between the two types of structure appears to be consistent with that found earlier by Deiterding and Bader [19].

Spinning motion of detonation was first observed in early experiments in the 1920's and was also reported by Schott [26]. In recent years, more studies have reported on this phenomenon [27-35], which can occur in both circular pipe and rectangular ducts [18, 25, 27-35]. Despite all these studies on spinning detonation, the dynamic behaviour and the mechanism of spinning detonation are still not completely clear. It appears that this type of detonation can only exist in narrow ducts the width of which is less than the width of the typical detonation cell size. (Henceforth, we shall describe a *narrow duct* as one with a width smaller than the typical detonation cell size consistent with the combustible gaseous fuel in view.) Zhang and Gronig [27] performed detonation experiments in a circular pipe for a two-phase system. They concluded that the transverse waves play a dominant role in stable detonation propagation in their two-phase system. In recent years, Tsuboi and co-workers [18, 25, 35] have conducted fairly extensive simulations on this issue for both the circular pipe and rectangular ducts. Even more recent work by Dou et al [23, 36] has shown the existence of spinning detonation in 3D simulations in narrow ducts, and an absence of spinning detonation front features in the larger ducts. They found that a self-sustained detonation will propagate in a narrow duct if the detonation front assumes a spinning mode. It should be mentioned that the size of calculated duct is much smaller than that decided by the criterion given by Vasil'ev [37]. Therefore, the mechanism of spinning detonation propagation and the conditions under which this phenomenon forms, as well as the relevance to the other two detonation structures, namely rectangular and diagonal, need to be further studied.

In this study, the influence of an initial disturbance on the detonation structure in a narrow duct is investigated by numerical simulation. The governing equations are the Euler equations, using the single step Arrhenius model for kinetics. Two different types of initial disturbances are introduced along the detonation front in 3D simulation and the evolution of the detonation front is observed following these disturbances. The motivation of this study stems mainly from an interest in the effect of initial disturbance on the type of the detonation structure ultimately obtained. A second motivation is curiosity about the mechanism that generates spinning detonation. Based on the simulation results, the mechanism of the spinning detonation and the qualitative effect of a disturbance on detonation patterns and on detonation stability are discussed.

**2. Physical Model**

The governing equations describing the fluid flow and the detonation propagation are the three-dimensional Euler equations with a source term representing the chemical reaction process. In conservative form, these are written as [23, 36]

$$\frac{\partial U}{\partial t}+\frac{\partial F}{\partial x}+\frac{\partial G}{\partial y}+\frac{\partial H}{\partial z}=S \tag{1}$$

where the conserved variable vector U, the flux vectors F, G, and H as well as the source vector ω are given, respectively, as

$$U=\begin{bmatrix}\rho\\ \rho u\\ \rho v\\ \rho w\\ E\\ \rho Y\end{bmatrix}\quad F=\begin{bmatrix}\rho u\\ \rho u^2+p\\ \rho uv\\ \rho uw\\ (E+p)u\\ \rho uY\end{bmatrix}\quad G=\begin{bmatrix}\rho v\\ \rho uv\\ \rho v^2+p\\ \rho vw\\ (E+p)v\\ \rho vY\end{bmatrix}\quad H=\begin{bmatrix}\rho w\\ \rho uw\\ \rho vw\\ \rho w^2+p\\ (E+p)w\\ \rho wY\end{bmatrix}\quad S=\begin{bmatrix}0\\ 0\\ 0\\ 0\\ 0\\ \omega\end{bmatrix}. \tag{2}$$

Here u, v, and w are the components of the fluid velocity in the x, y, and z directions, respectively, in a Cartesian coordinates system, $\rho$ is the density, p is the pressure, E is the total energy per unit volume, and Y is the mass fraction of the reactant. The total energy E is defined as

$$E=\frac{p}{\gamma-1}+\frac{1}{2}\rho(u^2+v^2+w^2)+\rho qY, \tag{3}$$

where $q$ is the heat production by the reaction, and $\gamma = Cp/Cv$ is the ratio of specific heats. Here, the constant pressure specific heat $Cp$ and constant volume specific heat $Cv$ are assumed to be constant, respectively. The source term $\omega$ is assumed to have an Arrhenius form

$$\omega = -K\rho Y e^{-(T_i/T)} \tag{4}$$

where $T$ is the temperature, $T_i$ is the activation temperature, and $K$ is a constant rate coefficient. For a perfect gas, the state equation is

$$p = \rho RT. \tag{5}$$

As such, equations (1) to (5) constitute a closed system of equations. This system of equations can be solved using a proper numerical technique.

The above mentioned equations are made dimensionless based on the state of the unburned gas,

$$\bar{\rho} = \frac{\rho}{\rho_0}, \bar{p} = \frac{p}{p_0}, \bar{T} = \frac{T}{T_0}, \bar{u} = \frac{u}{u_0}, \bar{v} = \frac{v}{u_0}, \bar{w} = \frac{w}{u_0}, \bar{x} = \frac{x}{x_0}, \bar{t} = \frac{t}{t_0}, \bar{E} = \frac{E}{p_0},$$

$$\bar{K} = \frac{Kx_0}{u_0}, \bar{q} = \frac{q}{u_0^2}, \bar{T}_i = \frac{T_i}{T_0},$$

where $u_0 = \sqrt{RT_0}$ and $t_0 = \frac{x_0}{u_0}$. The reference length $x_0$ is chosen as the half-reaction length (L), which is defined as the distance between the detonation front and the point where half of the reactant is consumed by chemical reaction in a ZND (Zel'dovich, von Neumann, and Doering) detonation. The overdrive parameter $f = (D/D_{cj})^2$, where $D$ is the detonation velocity and $D_{cj}$ is the Chapman-Jouguet (CJ) detonation velocity, which is calculated in closed form [23]. The reaction-rate pre-exponential factor K, sets the spatial and temporal scales.

Because of the self-similarity of the Euler equations, its dimensionless form and its original form are identical under the current scaling. For convenience, the overbar on each variable is dropped in the following sections. Here, the single step Arrhenius model is selected to reveal essential detonation physics not encumbered by complex chemical kinetics. In addition, many results are available in the literature for the single step model, which can be used for comparison. 2D simulations using more complex kinetics include work in references [12-13, 19].

The solution domain is a three-dimensional duct with square cross-section. Initial conditions consist of a planar ZND wave to which a perturbation (discussed in detail below) is superimposed. To maintain a specified overdrive f, support is provided by a piston located at infinity on the burnt side.

## 3. Numerical Implementation

The system of conservation laws of inviscid flow combined with the one-step chemical reaction model are discretized spatially in the space of eigenvectors in Cartesian coordinates using the fifth-order WENO (Weighted Essentially NonOscillatory) scheme [36, 38], while time discretization uses a 3rd order TVD (total variation diminishing) Runge-Kutta method.

The code for non-reactive Euler flow has been validated for the steady flow of oblique shock wave past a wedge, and for unsteady flow in one-dimensional shock tubes for both the so called Lax and Sod problems [38, 39]. For all the three examples above, the accuracy of the simulated results is good, as these results are in close agreement with analysis and other published work. Interested readers may refer to [23, 36] for details. Further validation was made for 1D simulation of detonation (by setting the mesh number be unity in the y and z directions) that was compared with analytical results.

In detonation computations the time step limitation is also affected by the reaction source term, in addition to the Euler convective terms. He and Karagozian [40] used the magnitude of the source term to determine the CFL condition. Here we also adopt the same approach to set the CFL condition, using a CFL value of 0.2 to 0.5 for 1D simulations and 0.1 to 0.2 in 3D.

The ZND solution used as initial condition is obtained directly from a separate one-dimensional simulation. The solution is captured in an inertial frame of reference moving at the steady ZND detonation velocity. As to the numerical boundary conditions, unburnt mixture is taken to enter the domain supersonically at a speed determined from the overdrive. Here, the theoretical CJ velocity normalized by $\sqrt{RT_0}$ is $u_{CJ}$ =6.809. The outflow boundary is set to be non-reflective since the flow behavior downstream the boundary does not affect the flow upstream. The walls are set to be reflective. The present treatments of boundary conditions are consistent and similar to well-established approaches adopted in previous studies [10-13, 23,36].

## 4. Results and Discussions

The case $f$=1.0, q=50, $T_i$ =20, and $\gamma$=1.20 is considered here; this case has been used in numerous previous numerical studies in one and two dimensions [20]. The computational geometry is a rectangular duct, and its size normalized by the reference length $L$ is 8x4x4. For the case used in present study, the width of the detonation cell is observed to be 20 times the reference length [23, 36]. Thus, the width of the duct is about 1/5 of the detonation cell width (as found in 2D simulations or 3D simulations in a much larger domain). The grid resolution was tested initially. A resolution of 121x61x61 grid points in the 8x4x4 domain, which corresponds to a resolution of 15 points in the ZND half reaction length, was found to be satisfactory.

In the initial conditions, first, a random 3D perturbation was added at the first time step, corresponding to a localized explosion just behind the leading shock. The form of the disturbance is $e^* = e + \alpha e g$. Here, $e^*$ is the perturbed total specific energy, including additional small fluctuations imposed on the reaction, g is a random number in a range from -1 to 1, and $\alpha$ is a coefficient, $0<\alpha<1.0$, that sets the amplitude of the fluctuation [13]. This random disturbance was imposed on the whole front cross-section. A second, symmetrical disturbance was tested, consisting in a random disturbance in a band, placed parallel to a diagonal of the square cross-section.

**(A) Flow patterns of 3D detonation with uniform random disturbance**

A random disturbance was introduced on the whole rectangular area of the detonation front at the beginning of the calculation t=0, and subsequently, the solution gradually evolved with time. Eventually, a high pressure zone arose at a corner and caused the flow pattern to become asymmetrical (Fig. 1). That region of high pressure did not stay in the corner, but it started moving around the four walls. Finally, after a long time, a sustained and fairly stable spinning detonation front appeared which consists of two waves along the two transverse directions respectively. In the present simulations in a narrow duct, the spinning front happened to take on a clockwise orientation, similarly to Tsuboi et al [25]. However, the rotating direction of spinning detonation may not matter; it is likely related to the distribution of the initial disturbance in the duct. As shown in our previous study [23, 36], the spinning motion of the detonation front appears only in narrow ducts. For wider ducts (with width close to the cell size), no spinning motion is observed. Obviously spinning detonations can not be appear in 2D simulations, in which case, for duct width smaller than the cell size, the detonation cannot be sustained.

Figure 2 shows the evolution of the spinning detonation front during one period. Two transverse waves are present, perpendicular to each other, propagating with the same period or frequency. This is illustrated schematically in Fig. 3. Figures 2 and 3 show that the spinning detonation is driven by two transverse waves moving in the two directions. Each collision of these two transverse waves on the walls generates high pressure and forms a hot spot which acts as a reaction center, periodically reigniting the wave hence ensuring its continued propagation. When this hot spot moves along the front area, it ignites the mixture, releasing energy. The two waves in the two directions on the cross section alternately control the reaction hence the detonation strength. The phase difference between these two waves determines the stability of the detonation. For a spinning detonation, there should be an optimum phase difference, sustaining the detonation. As the spinning detonation develops, it appears to stabilize be automatically at this phase difference.

Figures 2 and 3 indicate that the phase difference is close to 90°. In Fig. 3, the front is divided into 3 main regions by the two transverse waves, namely, Mach stem (MS), second Mach Stem (SMS), and incident shock (IS). Here, the region between the Mach stem and the incident shock is defined as second Mach stem (SMS). As time evolves, these three regions are transformed sequentially into one another, and every region undergoes an MS-SMS-IS transformation.

The generation of MS is produced by the collision of a transverse wave (TW) with a solid wall. For example, in Fig. 3, the horizontal TW collides with the upper wall -see Fig. 3 (b)- thereby leading to a MS in Fig. 3 (c); this MS extends downwards to cover about half the area of the front in Fig. 3 (d). Next, this MS becomes an SMS in Fig. 3 (e) and transforms to an IS as depicted in Fig. 3 (f). Next, the reflection at the lower wall gives rise to an MS travelling upwards, Fig. 3 (g), and the cycle repeats itself with Fig. 3 (a) and Fig. 3 (i) being identical. A similar sequence is identified for the vertical TW, starting from Fig. 3 (a). Figure 3 shows that the MS-SMS-IS direction corresponds to counter clockwise rotation, as indicated by thick arrows in Fig. 3 (g), while the motion of the spinning detonation is in clockwise direction, as seen from the evolution of the Mach stems in Fig. 3.

From the above, it is clear that Mach stems are generated by transverse waves alternatively colliding with the walls, supporting the spinning detonation in the rectangular duct. However, in spinning detonations in circular tubes, there is only one transverse wave, which rotates in one direction [25, 27]. Thus, it appears that there are differences between spinning detonations in these two geometries.

Figure 4 shows the cell patterns recorded on the side walls. The streamwise cell length is equal to approximately 19 reference lengths (the duct perimeter being equal to 16 reference lengths). The detonation track angle measured from the streamwise direction is close to 40 degrees. Experimental data for spinning detonation track angle, in both rectangular ducts and circular pipes have been reported in [25, 30, 35], with track angles measured from maximum pressure record ranging from 45 to 49 degrees. Simulation using detailed chemistry in [25] results in a track angle of 51 degree. As Tsuboi et al. discussed in [35], the magnitude of the track angle is mainly related to the mixture properties, and the specific heat ratio is the main parameter controlling the track angle. Thus it appears that simulation and experiment are reasonably consistent; a quantitative comparison would require simulations using the exact, correct mixture properties including chemical kinetics, which often remain uncertain.

There is certain degree of symmetry in the patterns on the four walls, so that patterns on opposite walls should show the same features. This is indeed observed from Figs. 2 and 3 and was also confirmed by simulations in [25]. Thus cell patterns are shown on only two walls in Fig. 4, on which it is observed that the transverse wave on one wall has a tail in each period besides a track, which was also noted by Tsuboi and Hayashi [25]. This behaviour will be explained from the flow patterns presented in Figs. 2 and 3 in the following paragraphs. Comparing the present simulation with those in [25], the reaction model employed is [25] is relatively more stable. As such, the record of the detonation cell pattern on the walls seems more regular and clear for the latter. However, the large scale features are clear and they are similar in both cases.

The spinning track is the record of the maximum pressure on the side walls as shown in Fig. 4. Its formation can be described as follow. Transverse waves (TW) produce equal strength records of maximum pressure on the two side walls perpendicular to this TW (Figs. 2 and 3). However, these records change when a TW moving in the orthogonal direction collides with these walls. In a period, while each TW crosses one full cell along the front in one direction, each of the TW moving in the orthogonal direction collides once on each of the two walls. In the first half of the period, the wall with which the collision occurs will produce a track due to the high pressure, while on the wall on which no collision occurs, a tail (a region of low pressure) appears. In the second-half of the period, the track and the tail are switched to be generated between the two walls.

**(B) Cells resulting from non-uniform symmetrical disturbance**

In this section, a random perturbation is only introduced on a banded area along a diagonal direction of the detonation front at the beginning of the calculation, at t=0. The width of the disturbed band width is one fourth of the duct width, and no disturbance is added on the remaining front area. As time increases, the amplitude of the disturbance grows symmetrically with respect to the diagonal. After a fairly long time, a periodic motion along the front is appears, that remains symmetrical with respect to the diagonal. Figure 5 shows the evolution of detonation front over one period. These patterns are schematically depicted in Fig. 6. Similar waves are observed in each two neighbouring wall. Figure 6 also shows that there are two transverse waves moving along the front diagonal and in the same direction. This mode is known as a "diagonal detonation." Figure 7 shows detailed contours of the flow and reaction variables corresponding to (e) of Fig. 5 and Fig. 6.

At later times, the initially symmetrical cell pattern becomes unstable and symmetry breaks down. The detonation weakens and is no longer self-sustained. As time continues increasing, the detonation strengthens again and it finally evolves into a spinning detonation (Figs. 8-11).

Figures 5 and 6 also shows two transverse waves on the front, propagating along diagonal directions and in phase. These two TWs generate one triple point on each wall. When a TW first reaches a corner, it is reflected and a new MS is produced. This new MS propagates back in the diagonal direction. Each of the two TWs move along one diagonal, and a symmetrical pattern is formed on the walls. Figure 6 shows the front divided into three regions by these transverse waves, namely, a Mach stem (MS), a second Mach Stem (SMS), and an incident shock (IS). Here, the region between the Mach stem and the incident shock is defined as second Mach stem (SMS). The MS and IS are transformed sequentially into one another; the SMS remains as such throughout the cell life cycle.

The MS drives the SMS and the SMS in turn drives the IS. This is fairly similar to the spinning detonation described in Fig. 3. However, here the driving motion occurs along a diagonal direction, while for the spinning detonation it occurs along a circumferential direction. In spinning detonations, MS, SMS and IS are transformed into each other sequentially, while in the symmetrical detonation, transformation can only occur between MS and IS.

Figures 8 to 11 show the various stages of the evolution as sequentially recorded along the walls, in the streamwise direction. Figure 8 shows the stage at which a symmetrical detonation pattern is formed. The initially uniform flow adopts this symmetrical pattern under the influence of the imposed symmetrical disturbance. As it propagates, the detonation first assumes a symmetrical pattern (Fig. 9), corresponding to the periodic diagonal mode. Figures 5 and 6 show identical patterns on each of the two walls. Next, on Fig. 10, the symmetrical disturbance breaks down. Finally, Fig. 11 shows formation of a self-sustained spinning detonation, which subsequently reaches a periodic regime identical to that shown in Fig. 4.

The symmetrical diagonal detonation in Figs. 5 to 7 differs from the diagonal mode reported in [22, 24], which corresponds to a much wider duct with at least of one cell width, in which case the detonation front is symmetrical with respect to both diagonals. In contrast, the detonation shown in Figs. 5 to 7 in a narrow duct is only symmetrical with respect to one diagonal. It is similar to one corner of the detonation discussed in [22, 24]. That the structures in Figs. 5 to 7 are self-sustained may be related to detonation stability issues.

**(C) Spinning detonation in a rectangular duct**

The detonation structure occurring in rectangular ducts can be classified as in-phase or out-of-phase. For duct width less than the cell width, there is no room for two triple points on any one of the side walls. Thus triple point collisions cannot occur on any wall.

If the two TWs on the neighbouring side walls are out-of-phase, the two TWs collide with the side walls at different times. Then, Mach stems, which correspond to hot spots, are generated alternatively by the two TWs. The strength of the detonation remains more or less constant, and the energy release does not suffer large fluctuations in time. Thus, the detonation is fairly stable. From these observations, it is noticed that the phase difference between the motion of the two

TWs results in spinning motion.

If the two orthogonal TWs are precisely in phase, spinning motion cannot appear and by symmetry, triple point collision can only occur at corners, leading to a symmetrical, diagonal detonation. Thus, whether the detonation adopts a diagonal structure depends upon the imposed disturbance. For the symmetrical diagonal detonation, the detonation strength experiences large oscillations (with time), and the energy release suffers large fluctuations. Therefore, in this case, the detonation is less stable.

For random disturbance, the detonation adopts an asymmetrical pattern from the beginning, which evolves relatively quickly into a spinning detonation. However, when the disturbance is symmetrically imposed, the detonation first evolves into a symmetrical, diagonal structure, which is unstable and cannot survive as such for a very long time. After a while, it decays and switches into an asymmetrical form, and it finally becomes a spinning detonation. Thus it appears that spinning detonation is stable in narrow ducts. One can thus attempt to control the detonation mode using the properties of the disturbance, to hasten appearance of a stable spinning detonation.

The results above are in agreement with experiments in a circular pipe [30]. Huang et al. concluded from calculations and experiments that the actual structure of the spinning detonation tries to match closely to the condition where the state parameters (pressure and temperature) reach their maximum values [30]. Thus the spinning detonation is the most stable for given mixture properties material, such that the energy release is maximized, as so are pressure and temperature (It is noticed that pressure and temperature represent total energy).

Maintaining constant self-sustained detonations may be useful in some industrial applications. As seen above, an initial random disturbance leads to a stable spinning detonation front much faster than non-uniform symmetrical disturbances. Thus the former is a more effective means to produce a stable detonation.

## 5. Conclusions

For the purpose of investigating the spinning detonation, the effect of different initial disturbances on detonations in narrow ducts was studied. Two types of disturbances were used to perturb the detonation front: uniformly random and banded symmetrical disturbances. For the random disturbance, a spinning detonation appears after some time. For the banded symmetrical disturbance, a symmetrical diagonal detonation is first formed. However, this diagonal detonation is unstable; it breaks down eventually and it finally turns again into a spinning detonation. The behavior of both disturbances leads to the conclusion that spinning detonation is the most stable mode. Thus, rapidly reaching a spinning mode appears to be the most effective way to produce a stable detonation in a narrow channel. This result may be useful if it is desired to trigger a deflagration to detonation transition process in a narrow duct. This confirms that attaining a sustainable detonation can be achieved by simply modulating the initial disturbance [24].

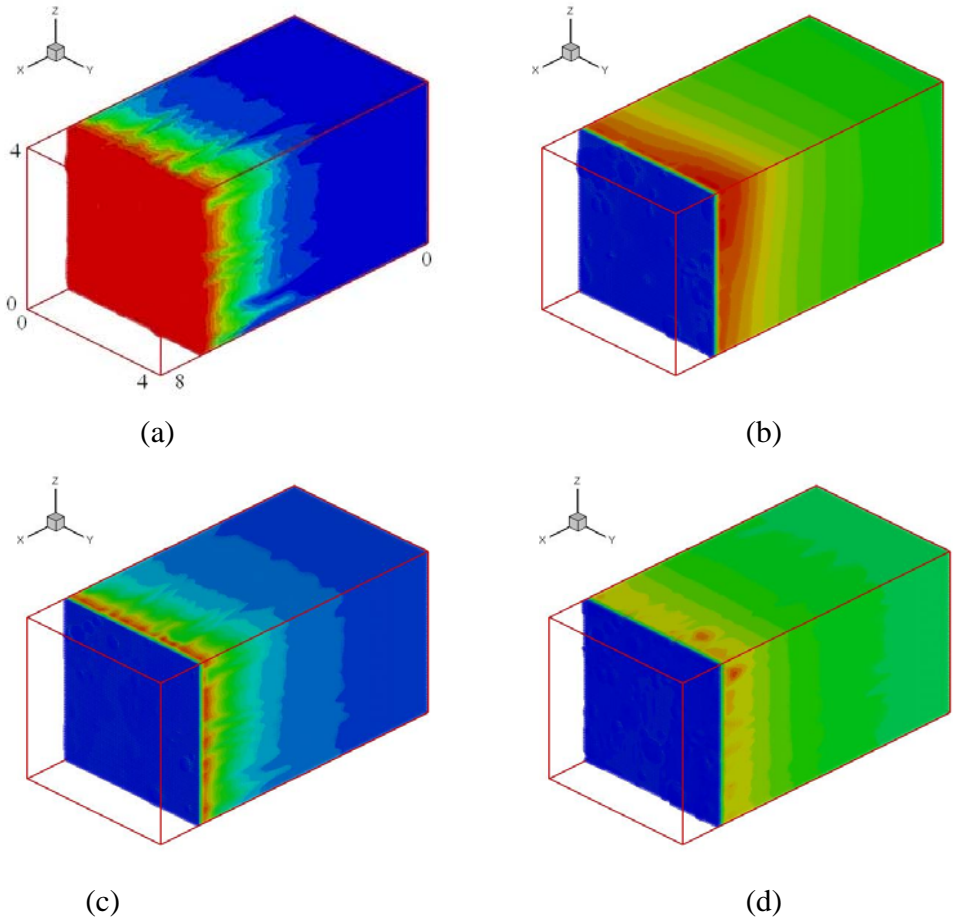

Fig. 1 Contours of the flow and reaction variables after 40,000 time steps (dimensionless time $t$=14.84), random disturbance. Parameters are $q$=50, $T_i$=20, $\gamma$=1.2, and $f$=1.0. (a) mass fraction of reactant; (b) pressure; (c) density; (d) streamwise velocity.

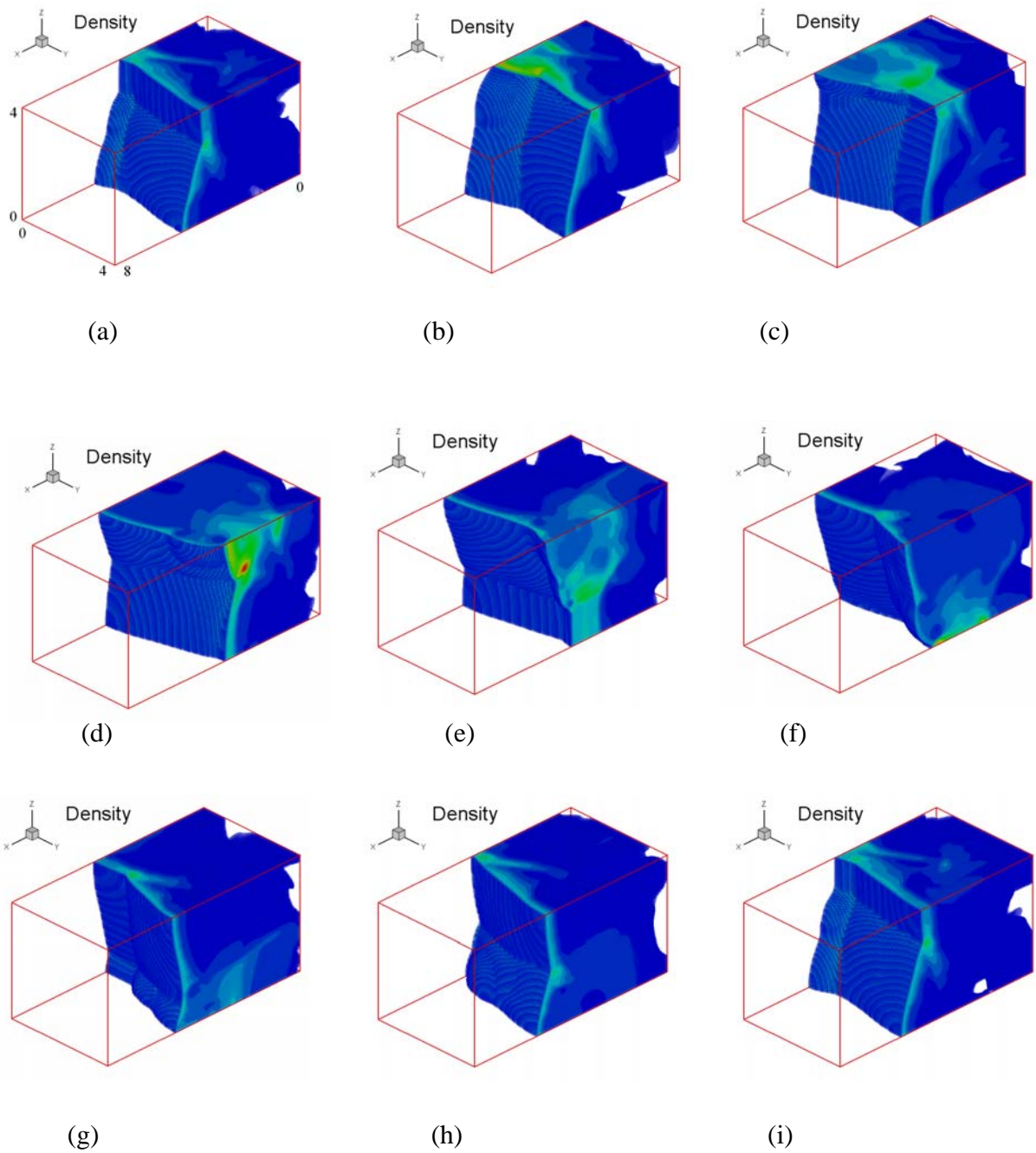

(a) (b) (c)

(d) (e) (f)

(g) (h) (i)

Fig. 2 Evolution of the detonation front on the walls over the full period (random disturbance). The flame front moves clockwise. Frames (a) – (i) are equally spaced in time. Time interval from (a) to (i) is 3.04.

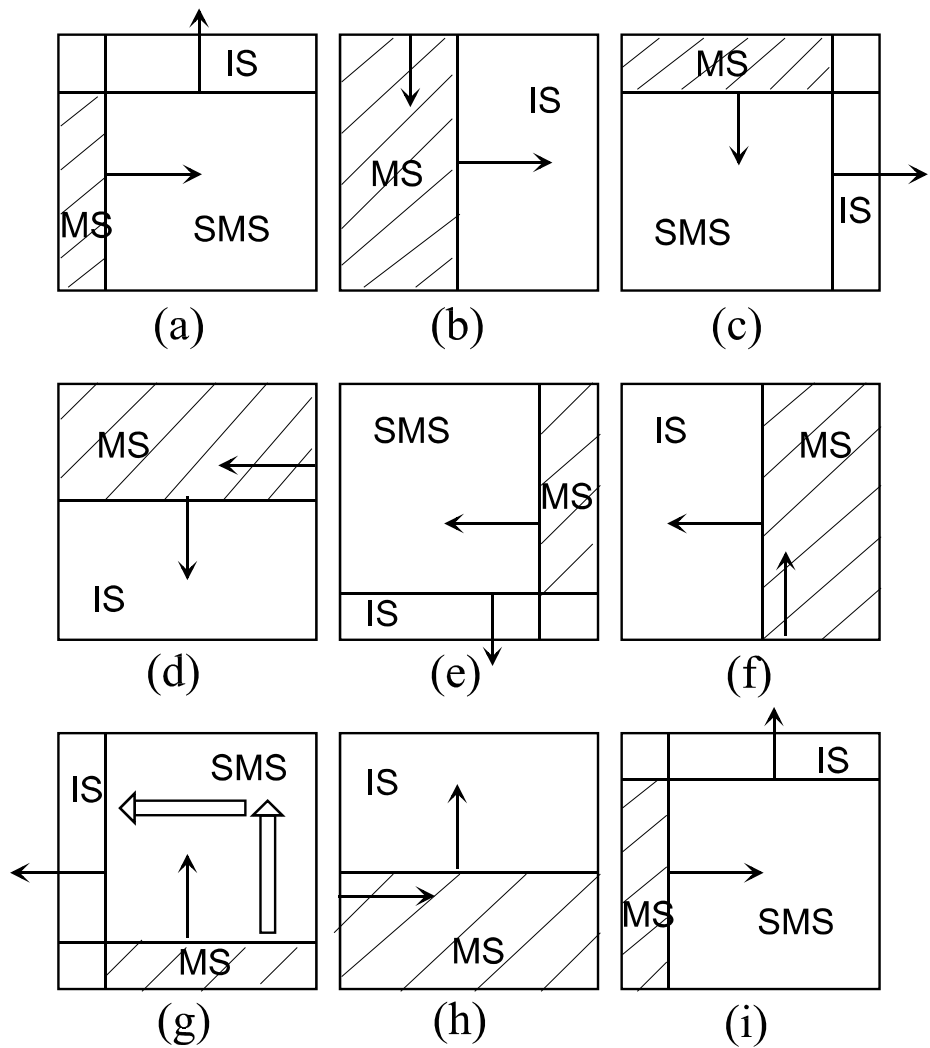

Fig. 3 Schematic front view of the detonation showing motion of the triple point lines over one period, for random disturbance. The transverse waves on the neighboring walls are out-of-phase. IS: incident shock; MS: Mach stem; SMS: Second Mach stem. The shaded areas are behind Mach stems. (a) to (i) corresponds respectively to (a) to (i) in Fig. 2.

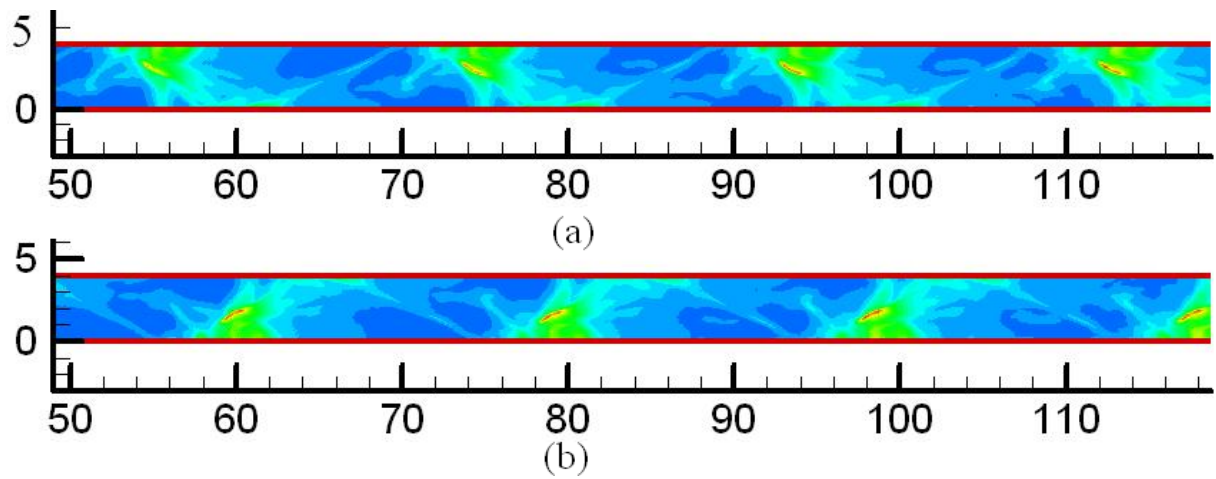

Fig. 4 History record of the maximum pressure on the walls, random disturbance (detonation propagates from left to right). (a) Side wall at z=0; (b) Side wall at y=0. The spinning angle measured from the streamwise direction is about 40 degree.

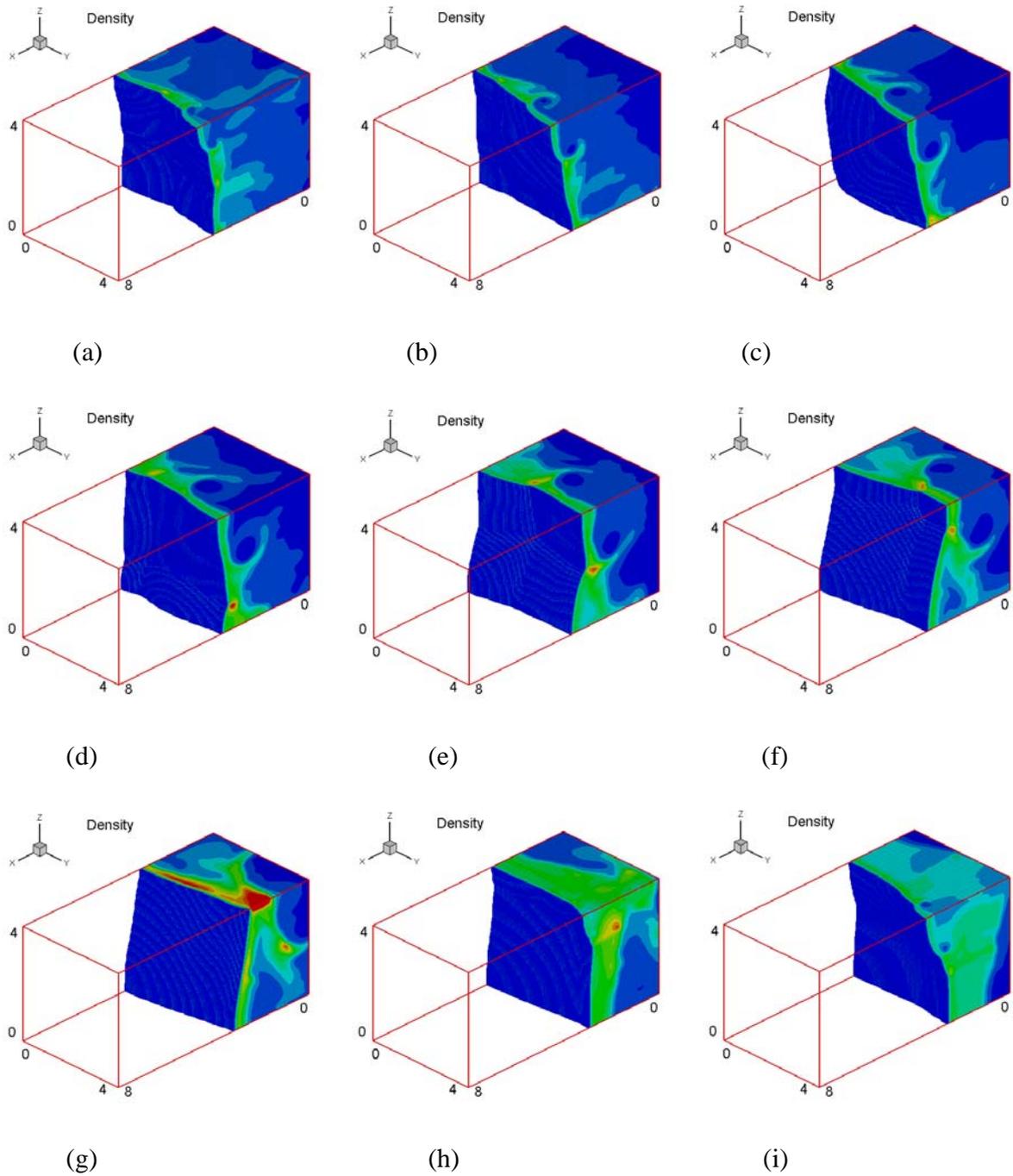

Fig. 5 Evolution of the detonation front on the walls for a full period, symmetrical disturbance: pressure evolution. The flame front moves along the diagonal direction. Frames (a) – (i) are equally spaced in time. Time interval from (a) to (i) is 3.23.

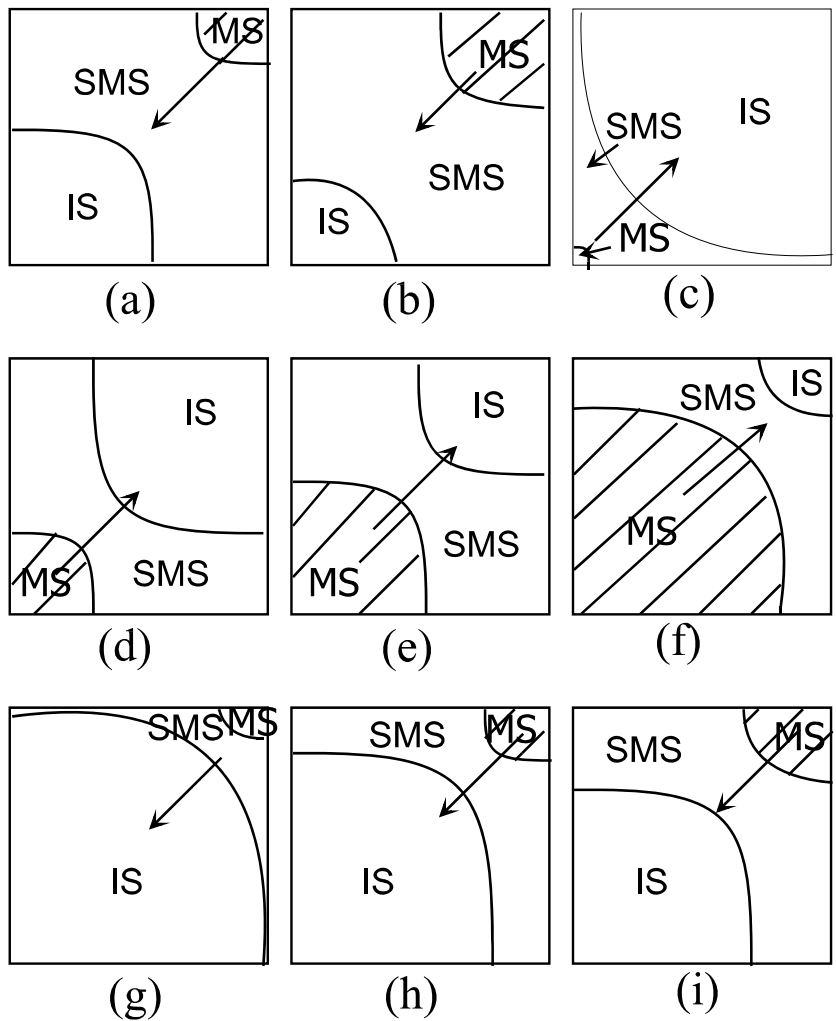

Fig. 6 Schematic front view of the detonation showing motion of the triple point line over one period, symmetrical detonation, from symmetrical disturbance. The transverse waves on each neighboring wall are in phase. The shaded areas are behind Mach stems. IS: incident shock; MS: Mach stem; SMS: Second Mach stem. (a) to (i) corresponds respectively to (a) to (i) in Fig. 5.

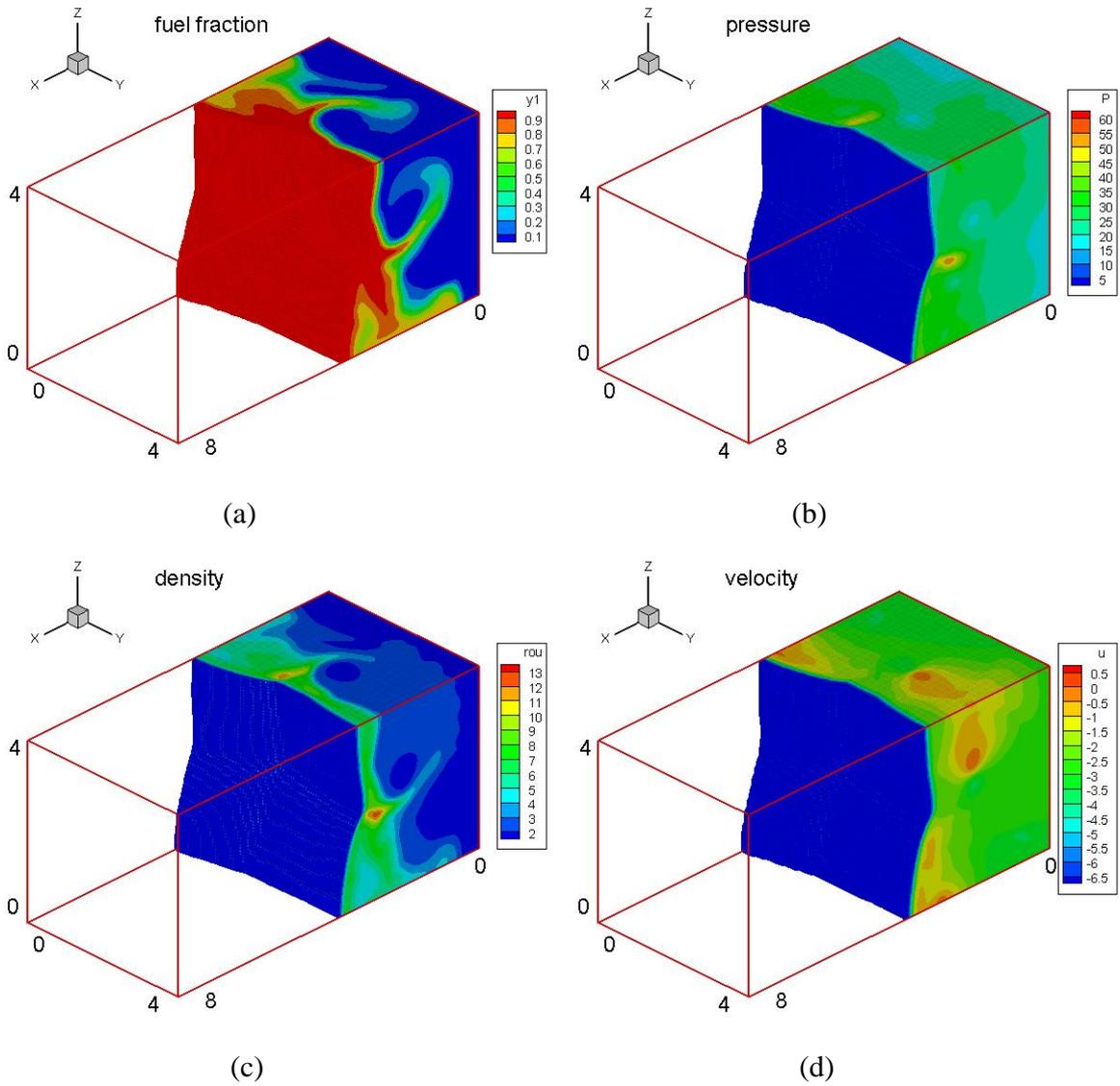

Fig. 7 Contours of the flow and reaction variables corresponding to (e) of Fig. 5, symmetrical disturbance. Parameters are $q$=50, $T_i$=20, $\gamma$=1.2, and $f$=1.0. (a) reactant mass fraction; (b) pressure; (c) density; (d) streamwise velocity.

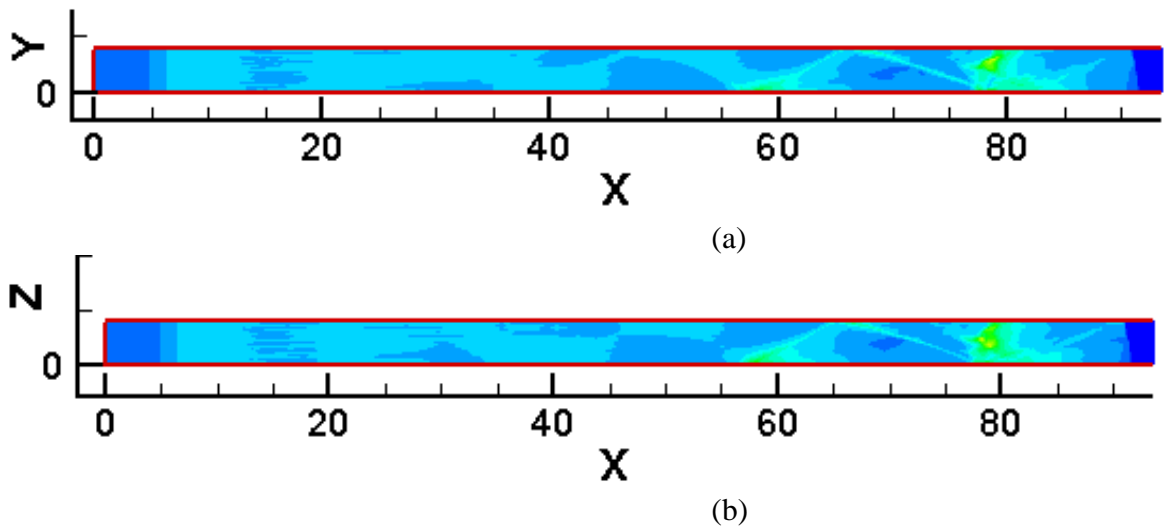

Fig. 8 History record of the maximum pressure on the walls, symmetrical disturbance (detonation propagates from left to right). (a) Side wall at z=0; (b) Side wall at y=0. Patterns on the two walls are symmetrical.

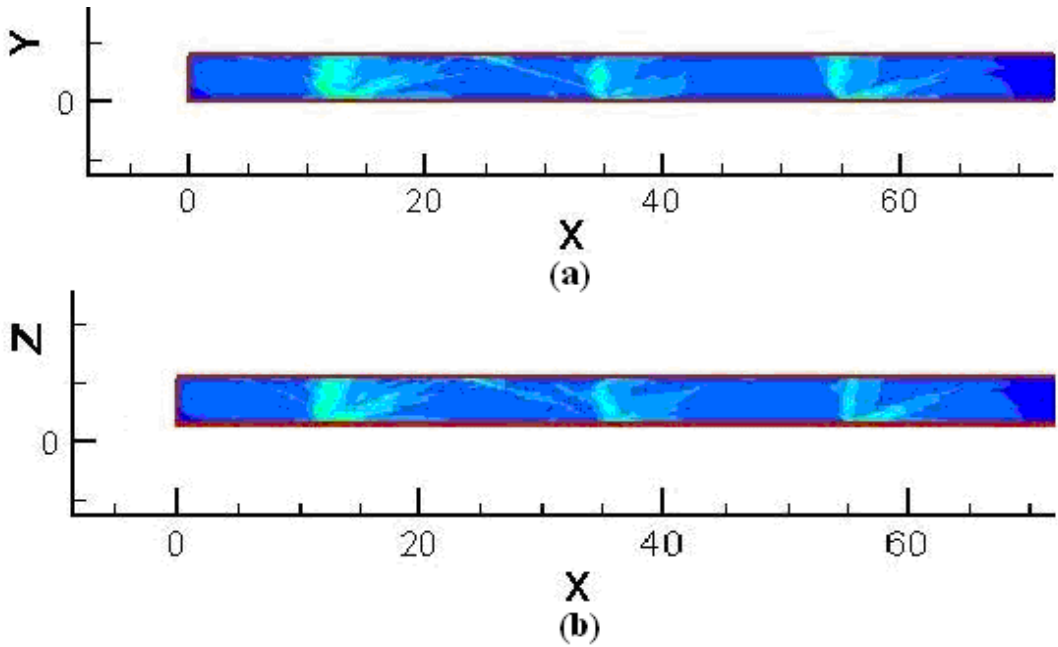

Fig. 9 History record of the maximum pressure on the walls, symmetrical disturbance (detonation propagates from left to right). (a) Side wall at z=0; (b) Side wall at y=0. Patterns on the two walls are the same. Periodic diagonal mode is formed. Transverse waves on neighboring walls are in phase. The number on the abscissa is only for the length scale. These patterns are downstream of those in Fig. 8.

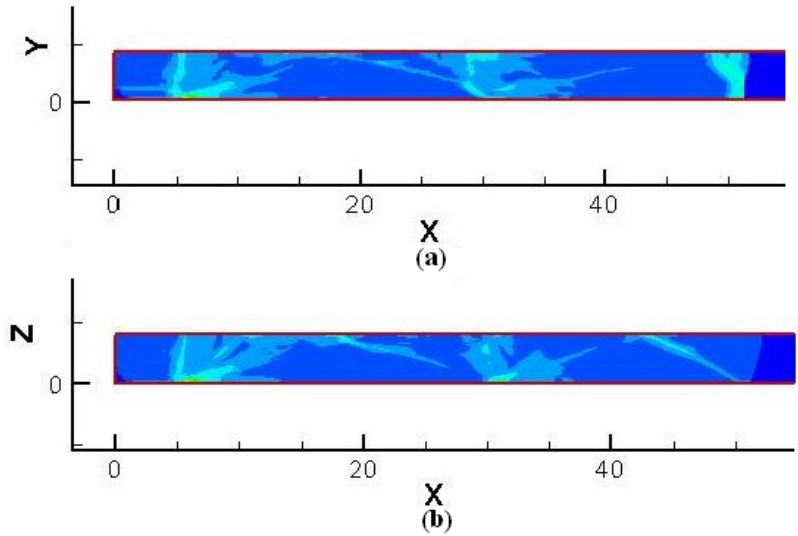

Fig. 10 History record of the maximum pressure on the walls, symmetrical disturbance (detonation propagates from left to right). (a) Side wall at z=0; (b) Side wall at y=0. The diagonal detonation mode decays, and symmetry is broken. Transverse waves on neighboring walls are out-of-phase. The number on the abscissa is only for the length scale. These patterns are downstream of those in Fig.9.

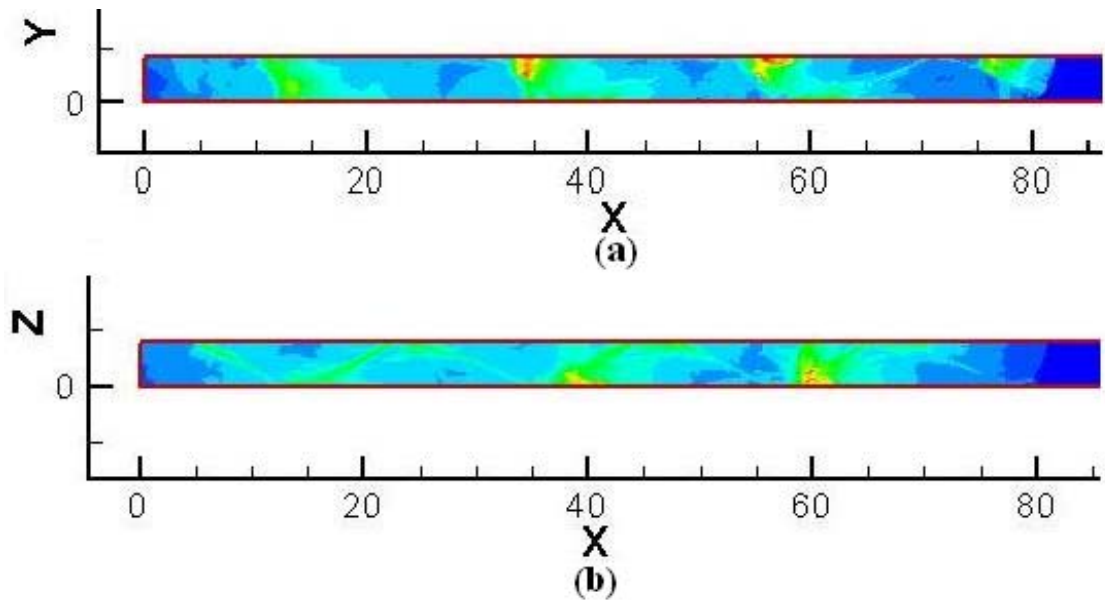

Fig.11 History record of the maximum pressure on the walls, symmetrical disturbance (detonation propagates from left to right). (a) Side wall at z=0; (b) Side wall at y=0. Patterns on the two walls evolve into a spinning detonation. Transverse waves on neighboring walls are out-of-phase. The number on the abscissa is only for the length scale. These patterns are downstream of those in Fig.10.